 \documentclass[showpacs,showkeys,amsfonts,amssymb,preprint]{revtex4}

\usepackage{amsmath}
\usepackage{graphics}
\usepackage[pdftex]{graphicx}
\usepackage{graphicx}
\usepackage{hyperref}

\usepackage[usenames,dvipsnames]{color}
\newcommand{\hide}[1]{}	%	for for personal notes

\newcommand{\Fs}[0]{F_\mathrm{s}}	%	for pre-stretching force
\newcommand{\Fd}[0]{F_\mathrm{d}}	%	for driving force
\newcommand{\Fshat}[0]{\hat{\Fs}}	%	for stretching force
\newcommand{\Fdhat}[0]{\hat{\Fd}}	%	for driving force
\newcommand{\kT}[0]{k_\mathrm{B}T}	%	for kBT

\begin{document}
\title{Reducing the variance in the translocation times by pre-stretching the polymer}
\author{Hendrick W. de Haan $^{1}$}
\author{David Sean $^{2,3}$}
\author{Gary W. Slater $^{2}$}
\affiliation{$^1$ Faculty of Science, University of Ontario Institute of Technology, Oshawa, Ontario, Canada, L1H 7K4}
\affiliation{$^2$ Physics Department, University of Ottawa, Ottawa, Ontario, Canada, K1N 6N5}
\affiliation{$^3$ Institut f\"ur Computerphysik, Universit\"at Stuttgart, 70569 Stuttgart, Germany}
\date{\today}

\begin{abstract}
Langevin Dynamics simulations of polymer translocation are performed where the polymer is stretched via two opposing forces applied on the first and last monomer before and during translocation.
In this setup, polymer translocation is achieved by imposing a bias between the two pulling forces such that there is net displacement towards the \textit{trans}-side.
Under the influence of stretching forces, the elongated polymer ensemble contains less variations in conformations compared to an unstretched ensemble.
Simulations demonstrate that this reduced spread in initial conformations yields a reduced variation in translocations times relative to the mean translocation time.
This effect is explored for different ratios of the amplitude of thermal fluctuations to driving forces to control for the relative influence of the thermal path sampled by the polymer.
Since the variance in translocation times is due to contributions coming from sampling both thermal noise and initial conformations, our simulations offer independent control over the two main sources of noise, and allow us to shed light on how they both contribute to translocation dynamics. 
Experimentally relevant conditions are highlighted and shown to correspond to a significant decrease in the spread of translocation times, thus indicating that stretching DNA prior to translocation could assist
in nanopore-based sequencing and sizing applications.
\end{abstract}

\maketitle

\section{Introduction}
The translocation of polymers across membranes through nanopores is central
in biological processes at the cellular level \cite{cell}
and in the development of DNA analysis techniques \cite{muthukumar2011}.
Applications such as nanopore-based sequencing technologies have led to 
numerous 
experimental \cite{ storm2005fast,chen2004probing,ivankin2014label,merchant2010dna,paik2012control,tahvildari2015integrating}, 
theoretical \cite{Sung1996,mondal2016stochastic,Saito2012,Saito:2011p7503,Sakaue:2007p7495,Sakaue:2010p7502,saito2013cis,Rowghanian:2011p7497,Rowghanian:2012}, and 
computational \cite{adhikari2015translocation,Suhonen2016,Menais2016,menaispolymer,2017arXiv170809184S,Sarabadani2017,Suhonen2017,Palyulin2014,Ikonen:2012p7489,Ikonen:2012p7492,huopaniemi2006langevin,Huopaniemi2007,forrey2007langevin,katkar2014effect,polson2015polymer} studies of polymer translocation.
Experimental histogram distributions of the translocation times $\tau$ exhibit surprisingly large variances --- particularly so when considering the perfect monodispersity of DNA.
This large variance in the translocation time is a prominent source of uncertainty
for applications such as using nanopores for DNA size determination.
It also introduces complications for sequencing applications where a consistent and orderly passage of bases is desirable.
Hence, techniques to reduce this variation are of great interest in the development of nanopore technologies.
In this work we propose and explore a methodology for reducing this variation by stretching the polymer
prior to translocation.

In an idealized translocation setup that does not include effects such as polymer-wall interactions, the degradation of the nanopore or interactions with ions or impurities in the solvent, 
the variance in translocation time is known to come from two different contributions: 
i) Brownian noise arising from thermal fluctuations
(i.e., the \textit{stochastic path} that a polymer follows in a particular translocation event)
and ii) conformational noise arising from otherwise identical polymers starting translocation in different conformations
\cite{lu2011origins,Saito2012,sarabadani2014,Sean2015,Sean2017}.
Hence, the variance in the translocation time can be reduced by either reducing the Brownian noise 
(e.g., turning down the temperature)
or by reducing the variation between the different initial conformations
(i.e., starting all translocation events from very similar initial states).

In this work, 
we narrow the distribution of initial conformations by stretching the polymer.
Prior to the translocation, the polymer is stretched by the application of a force to the first and last monomer in opposite directions.
As the stretching force is increased, conformations in which the polymer starts compressed near the pore
or even in a relaxed state become less and less likely.
In the limit of a very strong stretching force, 
all events would start with the polymer in a linear conformation along the pore axis.
Reducing conformational noise by confining the polymer in a long cylinder \cite{Sean2015} or a cylindrical cavity \cite{Sean2017} prior to translocation has been investigated previously.
However, the use of this dual-force stretching setup can reduce complications like knots and hairpins.

The stretching force, $\Fs$, thus enables us to implicitly control the range of initial conformations 
while Brownian fluctuations can be controlled via the simulation variable $\kT$.
Once the polymer is relaxed in a stretched state, 
the translocation process is initiated by increasing the force applied on the end of the polymer initially located in the pore by an amount $\Fd$ such that there is a net displacement towards the \textit{trans}-side through the nanopore.
Driving the translocation via end-pulling eliminates the need to consider the effects of monomer crowding on the \textit{trans}-side of the membrane, yielding a clean setup to investigate translocation dynamics.

\section{The stretching-pulling force setup}

\begin{figure}[h]
 	\centering
	\includegraphics[width=0.50\textwidth]{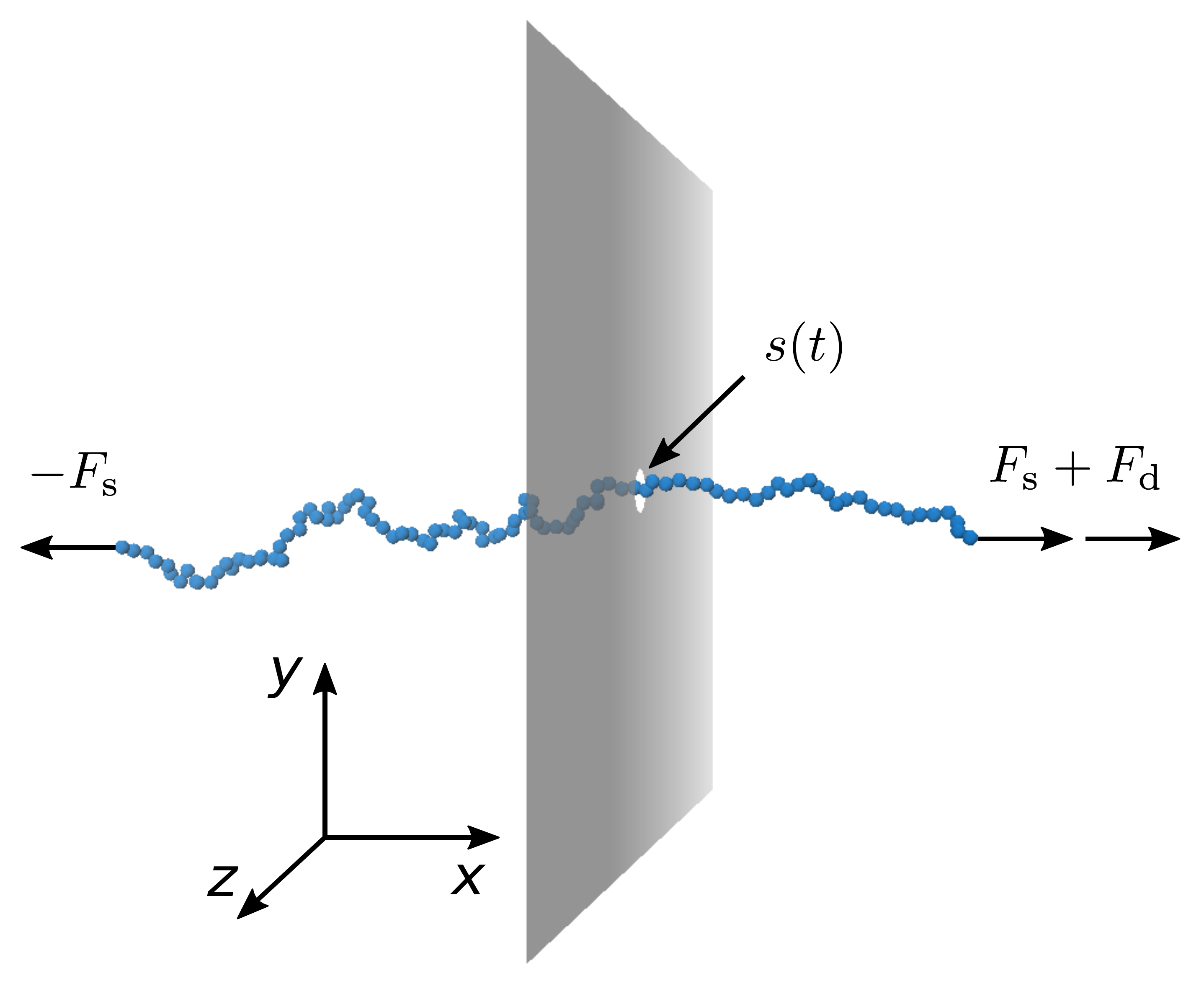} 
	\caption{(Color online) In this force configuration, the polymer is stretched by applying two equal and opposite forces $\pm \Fs \hat{x}$ on the polymer ends. An additional force $\Fd \hat{x}$ is applied to the first monomer to drive the polymer to the \textit{trans}-side. The translocation coordinate $s(t)$ is defined as the monomer index inside the pore.}
	\label{fig:scheme}
\end{figure}

Figure~\ref{fig:scheme} shows a schematic of how both a stretching force and a driving force are implemented in our simulations.
A stretching force $\Fs$ is applied to the first monomer ($i=1$) perpendicular to the plane of the membrane in the $trans$ direction while the same force is applied to the last monomer ($i=N$) in the opposite direction.
Although the force $\Fs$ stretches out the polymer, the net force on the polymer remains zero.
Hence, to drive the translocation process, an additional pulling force $\Fd$ is applied to the first monomer.
The DNA is prepared by stretching using $\Fs$ on opposite ends; these stretching forces continue to keep the DNA extended during the course of the translocation process.
This pulling force setup corresponds to translocation as controlled by optical tweezers or attached magnetic beads \cite{keyser2006optical,sischka2010dynamic,tabard2009single}.
While this has been previously studied via simulations \cite{Huopaniemi2007,2017arXiv170809184S}, the new feature here is the combination of the stretching and pulling forces.
A previous simulation study also employed a double force arrangement but in that work, a force in the pore opposed the pulling force \cite{Ollila2009}.

Lehtola et al. \cite{Lehtola2008} investigated how the initial polymer configuration affects the scaling law behaviors by simulating a polymer chain with an initial configuration of monomers in a straight line. They observed a scaling $\tau \sim N^{\alpha}$ with an exponent of $\alpha=2$, which is the same as the driven limit of Sung and Park \cite{Sung1996} where the monomer-fluid friction dominates over the monomer-nanopore friction. These findings indicate that, not only the applied force and the length of the chain, but also the initial polymer conformation affects the scaling exponent strongly.

\subsection{Polymer simulations}

A setup that is very close to a standard coarse-grained Langevin Dynamics (LD) simulation approach is used to model the system \cite{slater2009modeling}.
The excluded volume interactions between the $N=100$ monomer beads are implemented via the WCA potential \cite{weeks1978}:
\begin{eqnarray}
 U_\mathrm{WCA} (r) =
 \begin{cases} 
  3 \left(4 \epsilon_\mathrm{LJ} \left[ \left( \frac{\sigma}{r}\right)^{12} - \left( \frac{\sigma}{r} \right)^6 \right] + \epsilon  \right)       & \text{for } r  <  r_\mathrm{c}  \\
  0                                                                & \text{for } r \geq r_\mathrm{c}.
 \end{cases}
\label{WCA}
\end{eqnarray}
Here the nominal well depth $\epsilon_\mathrm{LJ}$ serves as a fundamental energy scale, and the nominal bead diameter $\sigma$ as the fundamental length scale.
Unless otherwise noted, we report length, energy and force variables in units of $\sigma$, $\epsilon_\mathrm{LJ}$, and $\epsilon_\mathrm{LJ} / \sigma$ respectively.
The cutoff distance is set to $r_\mathrm{c} =2^{1/6} \sigma$ such that this potential is purely repulsive.

To connect monomers along the contour length of the polymer, the finitely extensible nonlinear elastic (FENE) potential is used.
Note that for this work, as relatively high stretching forces are used, we multiply these two standard potentials (WCA and FENE) by a factor of three in order to reduce bond-stretching artifacts:
\begin{equation}
U_{\mathrm{FENE}}(r) = - 3 \left( \frac{1}{2} k r^2_0 \textrm{ln} \left( 1 - \frac{r^2}{r_0^2} \right) \right).
\end{equation}
To prevent bond crossing \cite{Grest:1986p4563} the parameters are chosen to be $k=30 \epsilon_\mathrm{LJ} / \sigma^2$ and $r_0=1.5 \sigma$.

Monomer positions are integrated in time using the Langevin equation of motion \cite{slater2009modeling}.
Monomers are subject to random thermal kicks which have a variance $2 \zeta k_\mathrm{B} T /\Delta t$ in each dimension, where $\Delta t=0.01$ is the integration time step.
Since we will vary the magnitude of thermal fluctuations via the variable of $\kT$, we report temporal quantities in units of $\zeta \sigma^2 / \epsilon_\mathrm{LJ}$.

The membrane is modelled as a mathematical surface with a pore of radius $1.5\sigma$.
The beads and the membrane interact via the WCA potential. 
The nanopore thus has an effective diameter of $2\sigma$ and a length of $\sigma$.

The external stretching force $\pm \Fs \hat{x}$ is applied on the first and last monomer (respectively) and the polymer is equilibrated.
The warmup procedure consists of pre-stretching the polymer with $-\Fs \hat{x}$ whilst having the first monomer confined in the pore.
After the warmup, the monomer held in the pore is released and pulled by additional external force $(\Fs+\Fd) \hat{x}$.
However, since the tension blob size scales as $\sim \kT/\Fs$, decreasing (increasing) the value of $\kT$ for a given stretching force $\Fs$ will yield an increased (decreased) end-to-end distance. 
To decouple this change in the amount of pre-stretching upon variations in the control parameter $\kT$, 
the polymer is stretched by pulling on both ends with a force $\Fs = \Fshat \kT / \sigma$.
This ensures that the dimensionless stretching force $\Fshat = \Fs \sigma / \kT$ yields the same degree of deformation.
Thus, when we report a particular value for $\Fshat$, this will always amount to the same degree of polymer deformation, irrespective of the value of $\kT$ used in the simulations.

The simulation is re-started with a new initial conformation whenever the polymer is found to be completely on the \textit{cis}-side of the membrane;
this is considered to be a failed translocation attempt.
A total of 1000 successful translocations are generated for each data point.

In a recent manuscript, we demonstrated that the dynamics of polymer translocation through a nanopore are significantly affected by
the balance between the magnitude of the force driving the polymer towards the \textit{trans}-side,
and the diffusive aspects arising from thermal noise \cite{haan2014}.
To quantify this relationship, one can define the \emph{translocation P\'{e}clet} number as the ratio between the polymer relaxation time and its translocation time $\tau$ \cite{Saito2012,dubbeldam2013driven,haan2014}:
\begin{equation}
P_t = \frac{1}{\tau} \frac{R_{g0}^2}{D_0} \approx  \frac{\Fd}{\kT},
\label{good_pec}
\end{equation}
where $R_{g0}$ and $D_0$ are the free solution radius of gyration and diffusion coefficient of the polymer respectively.
In that work, we found that typical coarse-grained (CG) simulation setups, where the force and diffusion coefficient are on the order of unity in LD units, yield a $P_t$ that is too low  for modelling realistic dsDNA translocation events \cite{haan2014}.
The two relevant energy scales are thus the thermal energy $\kT$ and the potential energy associated with monomers crossing the nanopore, $\Fd \sigma$, where $\sigma$ is the effective width of the membrane.
In the current simulations, we map the effect of pre-stretching over a wide range of $\Fd\sigma/\kT$ ratios --- from 1 to 100.
Values of interest are thus: $\Fd\sigma/kT \approx 1-10$, which corresponds to typical CG setups; $\Fd\sigma/\kT \approx 50$ which corresponds to a good (albeit too flexible) model of dsDNA; and $\Fd\sigma/\kT \geq 100$, which corresponds to even more coarse-grained models. 

In order to cover a range where $\Fd \in (1-100) \kT/\sigma$ while remaining both in the overdamped limit ($\Fd$ cannot be arbitrarily high) and being able to obtain events in a reasonable simulation time ($\Fd$ cannot be arbitrarily low), both $\Fd$ and $\kT$ are varied.
Simulations at driving forces of $\Fd \sigma /\epsilon_\mathrm{LJ}=0.1$, $0.2$, $0.5$, and $1.0$ are performed for two values of the thermal energy: $\kT=0.01 \epsilon_\mathrm{LJ}$ and $\kT=0.10 \epsilon_\mathrm{LJ}$.
To compare across cases, we report a dimensionless driving force $\Fdhat = \Fd \sigma/\kT$.
Note that the cases overlap at $\Fdhat=10$ --- attained using the two combinations: i) $\Fd=0.1 \epsilon_\mathrm{LJ}/\sigma$ with $\kT=0.01\epsilon_\mathrm{LJ}$; and ii) $\Fd=1\epsilon_\mathrm{LJ}/\sigma$ with $\kT=0.1\epsilon_\mathrm{LJ}$ --- which allows us to verify that there is only minimal differences between these cases
and that the effects overwhelmingly arise from the value of $\Fdhat$.

\section{Results}

\subsection{Translocation times $\tau$}
\begin{figure}[h]
 	\centering
	\includegraphics[width=0.5\textwidth]{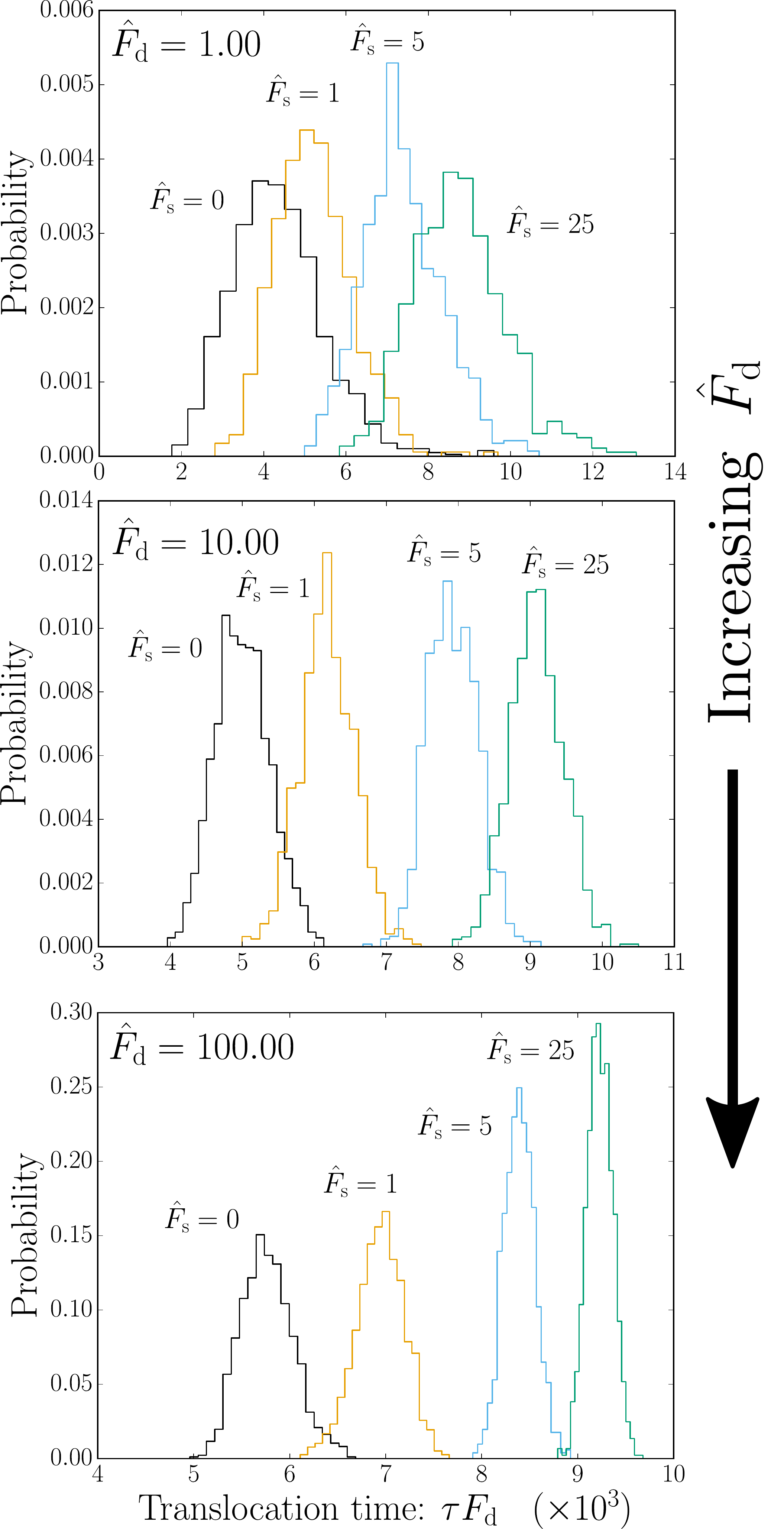} 
	\caption{(Color online) Histogram distributions of the translocation time with different stretching forces $\Fshat$. The three plots show different driving regimes ($\Fdhat=1$, $\Fdhat=10 $, and $\Fdhat=100$).}
	\label{fig:hist}
\end{figure}

Figure~\ref{fig:hist} shows the translocation time distributions at different stretching forces $\Fshat=0, 1, 5, 25$ for three driving forces $\Fdhat$.
Note that for these plots as well as others in the present manuscript, the translocation time $\tau$ is scaled by the driving force $\Fd$.
The absolute translocation time is trivially shorter for cases with a higher driving force;
the rescaling used here assumes that $\langle \tau \rangle \sim \Fd^{-1}$, which is not exactly the case, as we shall see later.
Examining the qualitative differences between the $\Fdhat$= 1, 10, 100 panels, one notes that the overlap between the distributions diminish with increasing driving magnitude $\Fdhat$.
Recall that the source of variance has contributions from both thermal noise and initial conformations; increasing $\Fdhat$ mostly reduces the contributions from the thermal noise, whereas increasing $\Fshat$ reduces the contributions from the initial conformations.
Thus the segregation between the $\Fshat$ histograms as $\Fdhat$ is increased arises from reducing the effect of thermal noise.
Conversely, the  $\Fdhat$=100 panel clearly showcases how conformational noise is suppressed with pre-stretching.
A striking feature here (and at all simulated $\Fdhat$ values) is that the mean translocation time $\langle \tau \rangle$ increases considerably with increasing $\Fshat$.

\begin{figure}[h]
 	\centering
	\includegraphics[width=0.60\textwidth]{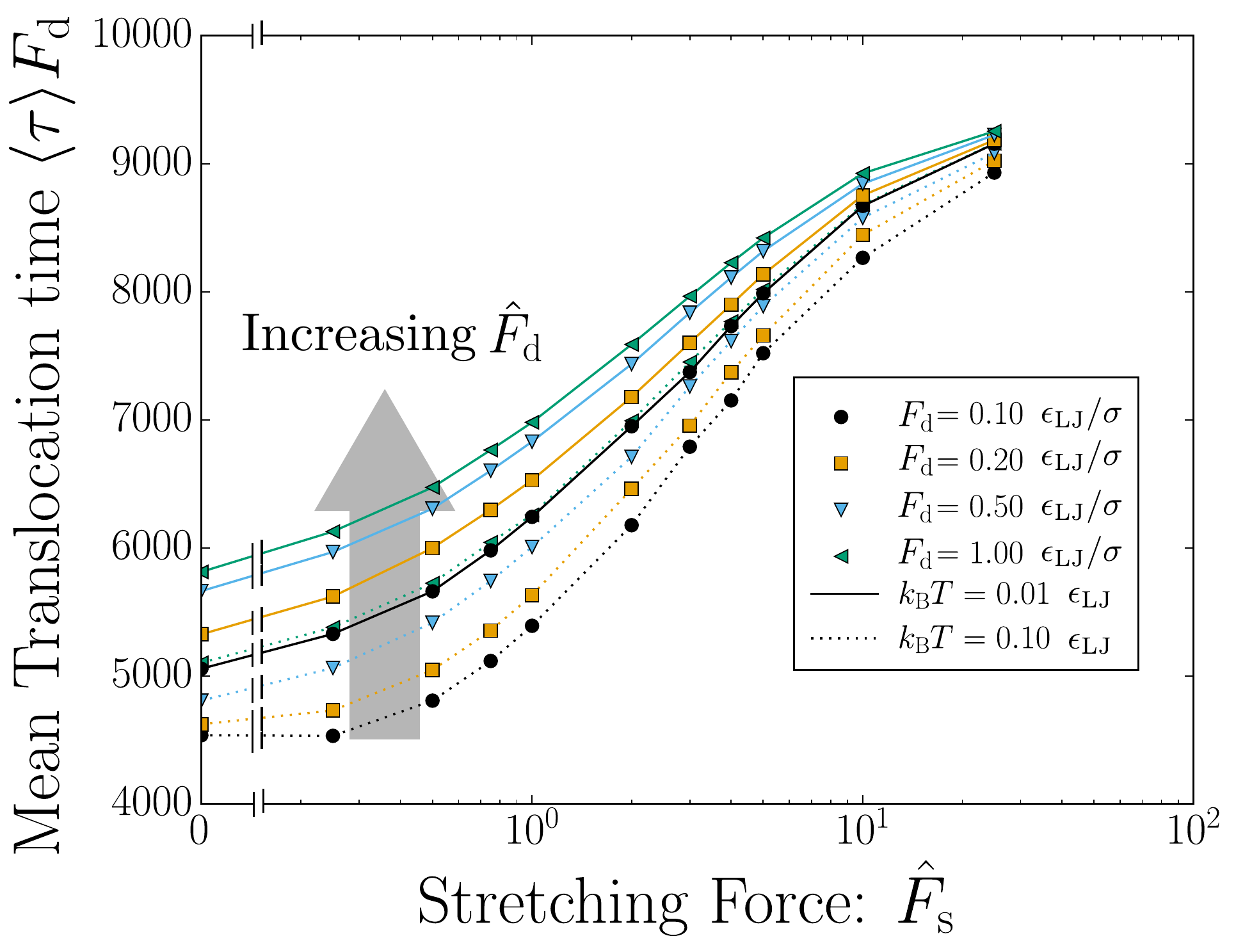} 
	\caption{(Color online) The average translocation $\langle \tau \rangle $ time scaled by the driving force $\Fd$ as a function of the scaled stretching force $\Fshat$.
}\label{fig:tau}
\end{figure}

This is shown more clearly in Fig. \ref{fig:tau} for all $\Fdhat$ values.
Again, the mean translocation time $\langle \tau \rangle$ is rescaled by the force $\Fd$.
The data in Fig.~\ref{fig:tau} exhibit two plateaus: one at low $\Fshat$ with no (or little) pre-stretching and one at $\Fshat \rightarrow \infty$ where the initial conformations are rod-like.
At $\Fshat=0$, the slight increase of $\langle \tau \rangle \Fd$ with increasing $\Fdhat$ highlights how the driving regime affects the translocation time.
Both datasets corresponding to $\Fdhat=10$ (with different combinations of $\kT$ and $\Fd$) give the same result, demonstrating that these simulations are indeed following the same the physical process.

In the high pre-stretching limit $\Fshat \rightarrow \infty$ the conformations all approach a rod-like initial state.
We expect the data to asymptotically collapse for the different $\Fdhat$ values here since the $\langle \tau \rangle \sim 1/\Fd$ should be strictly valid for rods, as confirmed in Fig.\ref{fig:tau}.

Two effects need to be highlighted in order to understand the increase in $\langle \tau \rangle$ with $\Fshat$.
First, we need to consider the tension-propagation (TP) translocation dynamics 
while taking into account the fact that tension propagates faster along a stretched polymer.
As the polymer is pulled through the pore the net force is simply the pulling force $\Fd$ applied to the first monomer.
However, the net drag on the polymer is proportional to the number of monomers that have been set into motion; monomers outside the range of the tension front do not contribute to the drag \cite{Saito2012,Saito:2011p7503,Sakaue:2007p7495,Sakaue:2010p7502,saito2013cis}.
When $\Fshat$ is low and the polymer is not in a stretched conformation,
the number of monomers being dragged increases as translocation proceeds,
but it is initially quite low as the pulling force is pulling out the slack in the coil rather than dragging the entire coil.
Conversely, when the stretching force is very high, the polymer starts nearly fully extended and 
thus more monomers contribute to the drag earlier;
the tension front then quickly reaches the end of the polymer at which point all monomers contribute to the drag.
This picture is consistent with the tension propagation model and has recently been explicitly tested for the case of driven polymer translocation from a tube \cite{Sean2015,Sean2017}.
The plateau observed at high $\Fshat$ indicates the saturation of the underlying polymer deformation.

For large $\Fdhat$ values, the driving force dominates over thermal forces,
hence the evolution of the polymer is largely deterministic:
the relative unimportance of thermal fluctuations means that the polymer does not relax as translocation progresses, i.e., the process is highly out of equilibrium.

Conversely, at small $\Fdhat$ values,
the polymer is allowed to partially adapt as translocation proceeds.
The driving force deforms the polymer via tension propagation which effectively moves the center of mass of the \textit{trans}-monomers \emph{away} from the nanopore.
If the polymer relaxes somewhat from this non-equlibrium state, it does so by primarily having monomers move \emph{towards} the pore.
Hence, as the thermal energy is increased, the effective drag is reduced and the translocation time decreases.
This is seen in Fig. \ref{fig:tau} at low $\Fshat$ 
where the higher $\kT$ curves are below the lower $\kT$ curves
and within each $\kT$ case, higher driving forces yield larger translocation times.
As the stretching force $\Fshat$ increases, the impact of the non-equilibrium effects is diminished.
At high $\Fshat$ values, the chain is nearly completely stretched and acts almost as a single body for the entire translocation process regardless of the balance between $\Fd$ and $\kT$.
Consequently, very little deviation of $\tau$ between $\Fdhat$ values is observed in this limit and the data converge.
Note that the $\Fshat=0$ case is studied in detail in ref \cite{haan2014} where we demonstrate that 
this additional increase in $\tau$ at low $\Fshat$ reflects varying degrees of non-equilibrium effects.

%%%%%%%%%%%%%%%%%%%%%%%%%%%%%
\subsection{Translocation time standard deviations $\sigma_\tau$}

Returning to Fig. \ref{fig:hist}, the variation in the widths $\sigma_\tau$ of the distributions depend not only on $\Fshat$ but also on $\Fdhat$.
In the last panel with $\Fdhat=100$, corresponding to the driving force dominating over thermal forces,
the width $\sigma_\tau$ decreases with increasing $\Fshat$.
For $\Fdhat=1$ (first panel) and $\Fdhat=10$ (middle panel), 
the behaviour is less obvious and the width is only weakly dependent upon $\Fshat$.
Comparing across panels, $\sigma_\tau$ decreases with increasing $\Fdhat$ at any particular $\Fshat$ value. 
This reflects the suppression of diffusion resulting in reduced variation of the stochastic paths and narrower distributions.

\begin{figure}[h]
 	\centering
	\includegraphics[width=0.60\textwidth]{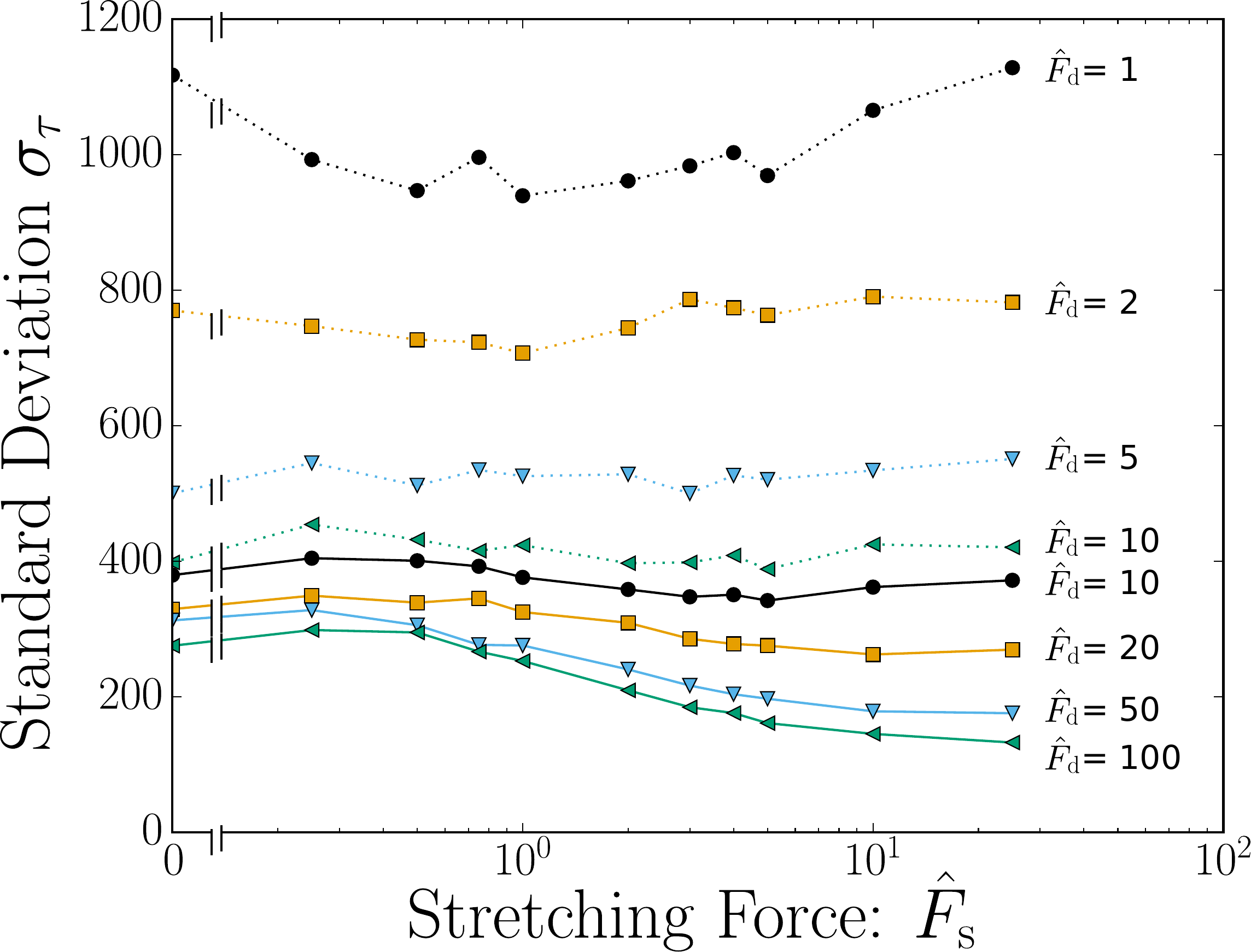} 
	\caption{(Color online) Standard Deviation $\sigma_\tau$ plotted as a function of the stretching force $\Fshat$ for several values of the driving force $\Fdhat$.}
	\label{fig:std}
\end{figure}

The value of $\sigma_\tau$ is plotted against $\Fshat$ for different $\Fdhat$ values in Fig. \ref{fig:std}.
As expected, $\sigma_\tau$ decreases with $\Fshat$ at large $\Fdhat$ values.
At $\Fdhat=1$, the width intially decreases but subsequently increases
while at $\Fdhat=2$ and 5, $\sigma_\tau$ is essentially independent of $\Fshat$.

\begin{figure}[h]
 	\centering
	\includegraphics[width=0.60\textwidth]{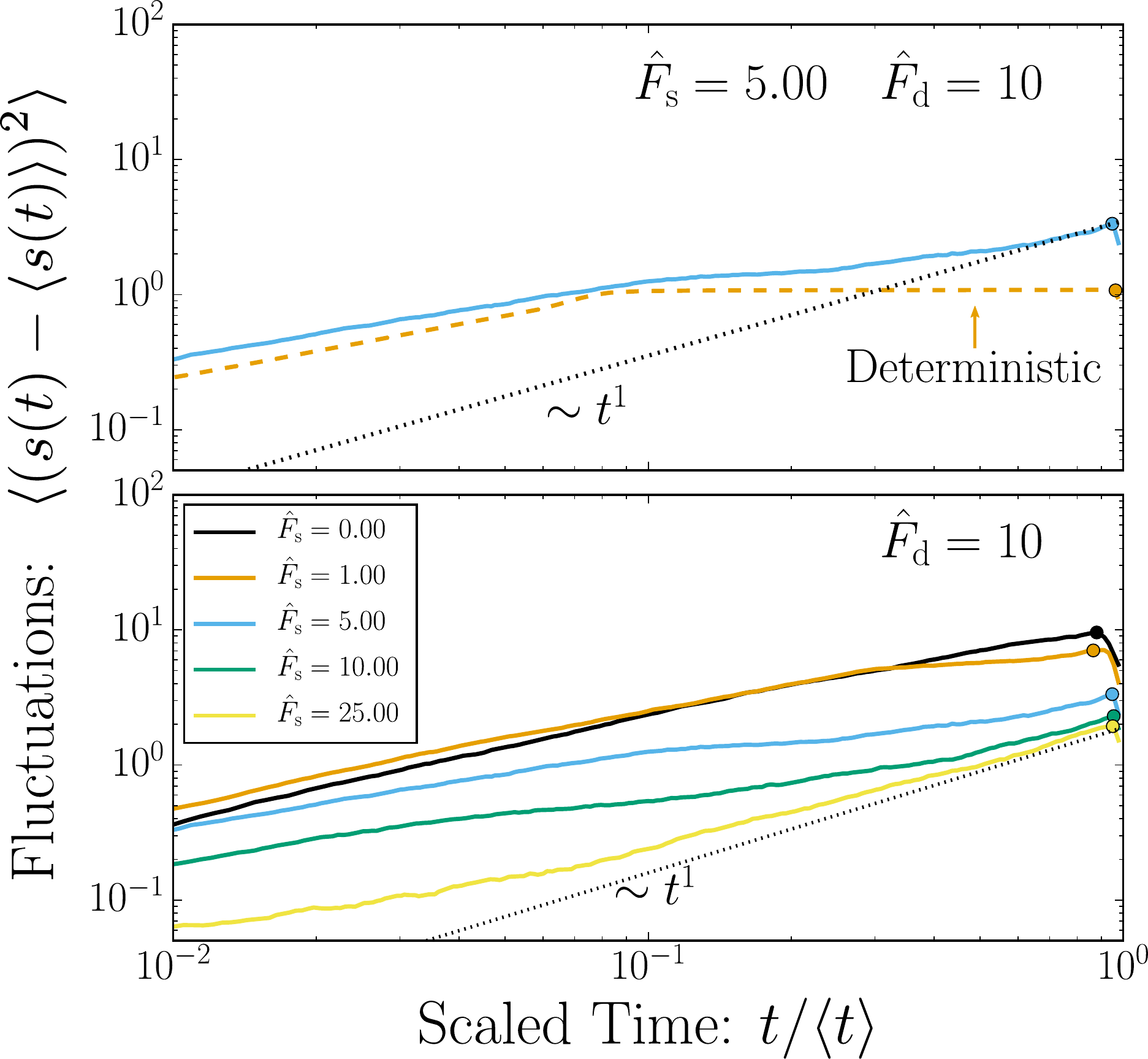} 
	\caption{(Color online) Fluctuations of $s(t)$ vs scaled time $t/\langle \tau \rangle$. The top panel shows a single stretching force together with a deterministic simulation and a $t^1$ line to guide the eye.
	The bottom panel shows different stretching situations as different curves. The circle symbol dot near $t=\langle \tau \rangle$ indicates the time for the first translocation of the ensemble, i.e., the ensemble population decreases from this point onwards.}
	\label{fig:stdb}
\end{figure}

As discussed, there are two contributing factors to the width of the distribution of translocation times: the ensemble of initial conformations and the effects of thermal noise  during the translocation process.
One expects that the latter mechanism, the variation due to diffusion, to grow with time.
Recalling that the translocation time increases significantly as the stretching force increases, this means that there are two competing effects for the data shown in Fig.~ \ref{fig:std}.
Due to the fact that the mean translocation time $\langle \tau \rangle$ increases with $\Fshat$, diffusion will obviously cause $\sigma_\tau$ to increase with $\Fshat$.
On the other hand, increasing $\Fshat$ reduces the ensemble of initial conformations and this causes a reduction in $\sigma_\tau$ with increasing $\Fshat$.

The competition between these effects yields the behaviour seen in Fig. \ref{fig:std}
At low $\Fdhat$ and low $\Fshat$, $\sigma_\tau$ is relatively independent of $\Fshat$ 
(excluding the non-monotonic case of $\Fdhat=1$)
indicating that these two effects almost balance out.
This is largely true also for the case of high $\Fdhat$ and low $\Fshat$.

However, the behaviour at high $\Fshat$ depends on $\Fdhat$.
At low $\Fdhat$, $\sigma_\tau$ increases with increasing $\Fshat$ indicating that the diffusive effects dominate over the further reduction in the range of initial conformations.
On the other hand, at high $\Fdhat$, $\sigma_\tau$ decreases with increasing $\Fshat$ and hence diffusive effects are marginal compared to the narrowing of the distributions that arises from the reduced conformational phase space.
Low $\Fdhat$ thus corresponds to diffusion dominated dynamics while for high $\Fdhat$ the process is primarily driven, as one would expect.

Up to this point, we have been interested in the fluctuations at a specific time, namely the end of the translocation process where the translocation coordinate $s(t)$ equals $N$ and the time $t$ equals $\tau$, $s(\tau)=N$.
Thus, the focus has been on the accumulated noise $\sigma_\tau$.
However, the physical picture put forth by the tension-propagation theory suggests that when an end-pulling process is considered \cite{Saito2012},
the noise arising from the initial conformations plateaus when the tension front reaches the last monomer.
After this stage the $N$ monomers of the polymer move in unison with a constant velocity, and the source of fluctuations is solely Brownian. 
Thus the way by which the fluctuations grow should exhibit the characteristic two-steps of the underlying dynamics.
To explore this, we examine statistics of the translocation coordinate $s$ as a function of time $t$.

Log-log plots of the variance $\sigma_s^2= \langle (s(t)-\langle s(t)\rangle )^2 
\rangle $ as a function of the scaled time $t/\langle \tau \rangle$ are shown in 
Fig.~\ref{fig:stdb}.
In the top panel of Fig.~\ref{fig:stdb} we show a selected $\Fshat=5$ case with two additional curves chosen to highlight the contrast between fluctuations arising from Brownian noise \textit{versus} those due to initial conformations.
To do this, we perform a set of deterministic simulations where the polymer is initiated with conformations according to $\Fshat=5$
but $\kT$ is set to zero such that there is no thermal noise.
Hence, the resulting fluctuations arise solely due to variations in the initial conditions.
These deterministic data are shown as a labeled dashed line.
To contrast this, we also plot the expected fluctuations arising from Brownian motion, modelled as a line with slope unity ($\sim D t^1$) with an arbitrary prefactor chosen to match the data at the end of the process.
This is shown as the dotted line in the top panel of Fig.~\ref{fig:stdb} (note that this line should be taken to guide the eye, as the instantaneous diffusion coefficient of the polymer depends upon  the fractional amount of monomers set in motion by the applied force).
%This would result in a time-dependent coefficient $D(t)$ which becomes constant only when the tension front has arrived at the last monomer. 
Examining the top panel of Fig.~\ref{fig:stdb} using these two additional curves, the fluctuations from the selected $\Fshat=5$ case start very near the deterministic result.
However, the deterministic line saturates near $t/ \langle t \rangle \approx 0.08$.
This corresponds to the time where the tension front reaches the last monomer.
Correspondingly, the fluctuations in the simulations begin to level off---but cannot completely flatten due to diffusion effects.
In general, it is assumed that the motion of a translocating polymer follows a power law $\langle \Delta s(t)^2  \rangle \sim t^\beta$ with $\beta$ close to unity \cite{dubbeldam2007driven}.
Fluctuations arising from thermal noise should eventually dominate over the effect from initial conformations which saturate before the translocation time $ \langle \tau \rangle$.
The proposed linear $\langle \Delta (t)^2  \rangle \sim t^1$ line closely follows the data for long times.

The family of curves presented in the bottom panel of Fig.~\ref{fig:stdb} represents data for varying stretching forces.
As expected, polymers with higher pre-stretching exhibit a lower amount of fluctuations: as $\Fshat$ is increased the curves shift downwards.
The curves differ not only in the absolute amount of fluctuations but also in the rate at which these fluctuations grow.
As expected in the final---Brownian noise only---stage of translocation, the fluctuations appear to follow the expected diffusive power law.
In situations of low pre-stretching, this diffusive regime is short-lived since the onset occurs very close to the end of the translocation process.
In the opposing limit of high pre-stretching, the diffusive regime completely dominates.
In early times however, fluctuations arise from both the conformational noise (molecular individualism) and Brownian diffusion.
Hence, pre-stretched polymers start with a low variance in $s(t)$ that increases approximately linearly with $t$
while polymers at $\Fshat=0$ start with a large variance in $s(t)$ and essentially does not exhibit the diffusive regime.
In between these extremes, a cross-over from initial conformation to thermal noise can be observed.
The crossover (close to the tension-propagation time) depends upon the amount of pre-stretching, consistent with references \cite{Sean2015,Rowghanian2012}.

%%%%%%%%%%%%%%%%%%%%%%%%%
\subsection{Scaled variations}

\begin{figure}[]
 	\centering
	\includegraphics[width=0.9\textwidth]{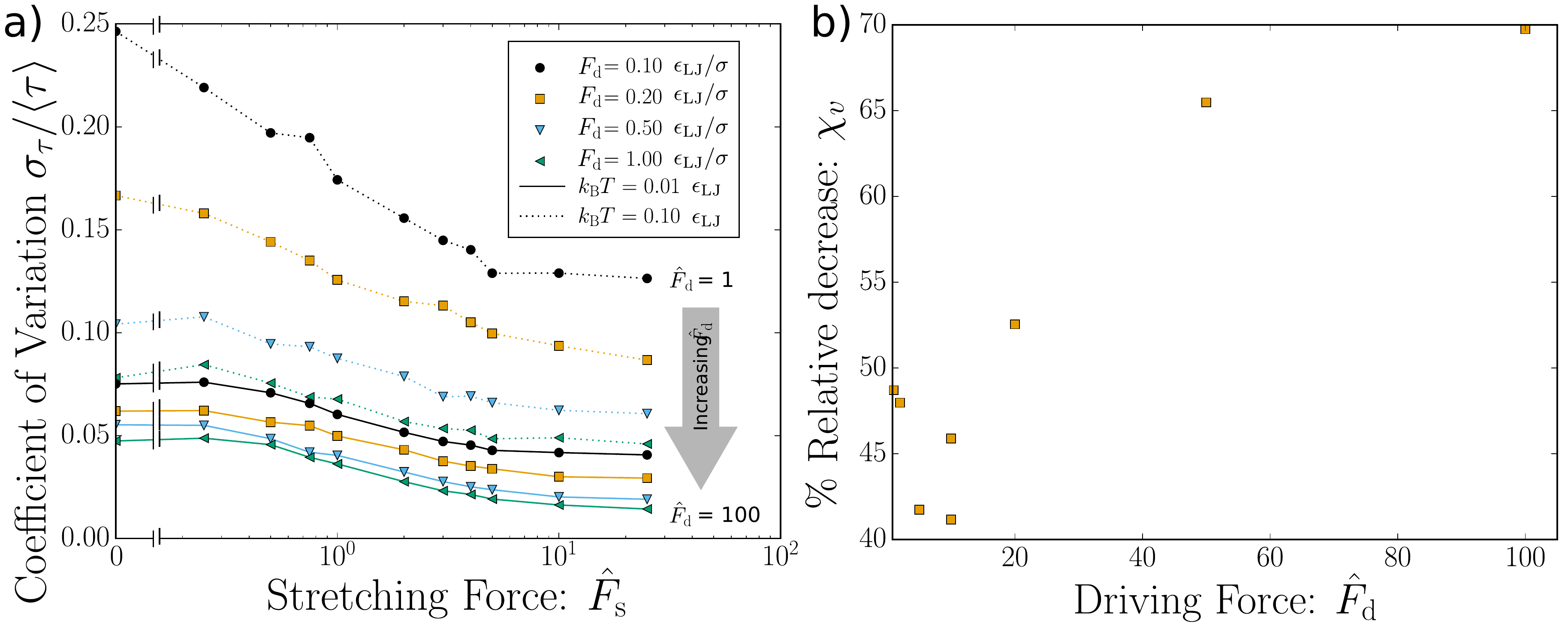} 
	\caption{(Color online) a) Coefficient of variation  $c_v=\sigma_\tau/\langle \tau \rangle$ plotted as a function of the stretching force $\Fshat$ for several values of the driving force $\Fdhat$. b) Plot of the  percent decrease of the coefficient of variation $c_v$ as a function of the driving force $\Fdhat$.}
	\label{fig:fluct}
\end{figure}

It is convenient to calculate the standard deviation $\sigma_\tau$ normalized by the mean translocation time $\langle \tau \rangle$.
This quantity, which reflects the relative width of the distributions, is known as the coefficient of variation and is given by
\begin{equation}
c_v = \frac{\sigma_\tau}{ \langle \tau \rangle}.
\end{equation}
These values are plotted in Fig. \ref{fig:fluct}a.
Now, almost without exception, the normalized distribution width decreases with increasing $\Fshat$
thus indicating that reducing the range of initial conformations does reduce the uncertainty in the measured translocation time, even though translocation takes longer.

To quantify these effects further, we plot in Fig~\ref{fig:fluct}b the percent difference in the coefficient of variation $\chi_v$.
It is calculated as the difference between $c_v$ with no stretching force and $c_v$ at the highest stretching force normalized by $c_v$ at $\Fshat = 0$:
\begin{equation}
\chi_v \equiv \frac{c_v ( \Fshat = 0 ) - c_v ( \Fshat = 25 )}{c_v ( \Fshat = 0 )} \times 100.
\end{equation}

The decrease in the variance of the translocation times is strongly dependent on the driving regime $\Fdhat$. 
For $\Fdhat = 1-10$ for instance, the percent relative decrease lies between $40 - 50$\%.
At these low driving forces, the relative decrease actually decreases slightly with increasing driving force.
Note that there is a slight discrepancy between the $\kT=0.1$ and $\kT=0.01$ cases which do not quite overlap at $\Fdhat=10$.
Above $\Fdhat = 10$, the relative decrease increases significantly with increased driving force. 
For $\Fdhat = 100$, the relative width of the distribution decreases by 70\% between $\Fshat = 0$ and $\Fshat = 25$. 
The data is saturating at large stretching forces.

\begin{figure}[h]
 	\centering
	\includegraphics[width=0.50\textwidth]{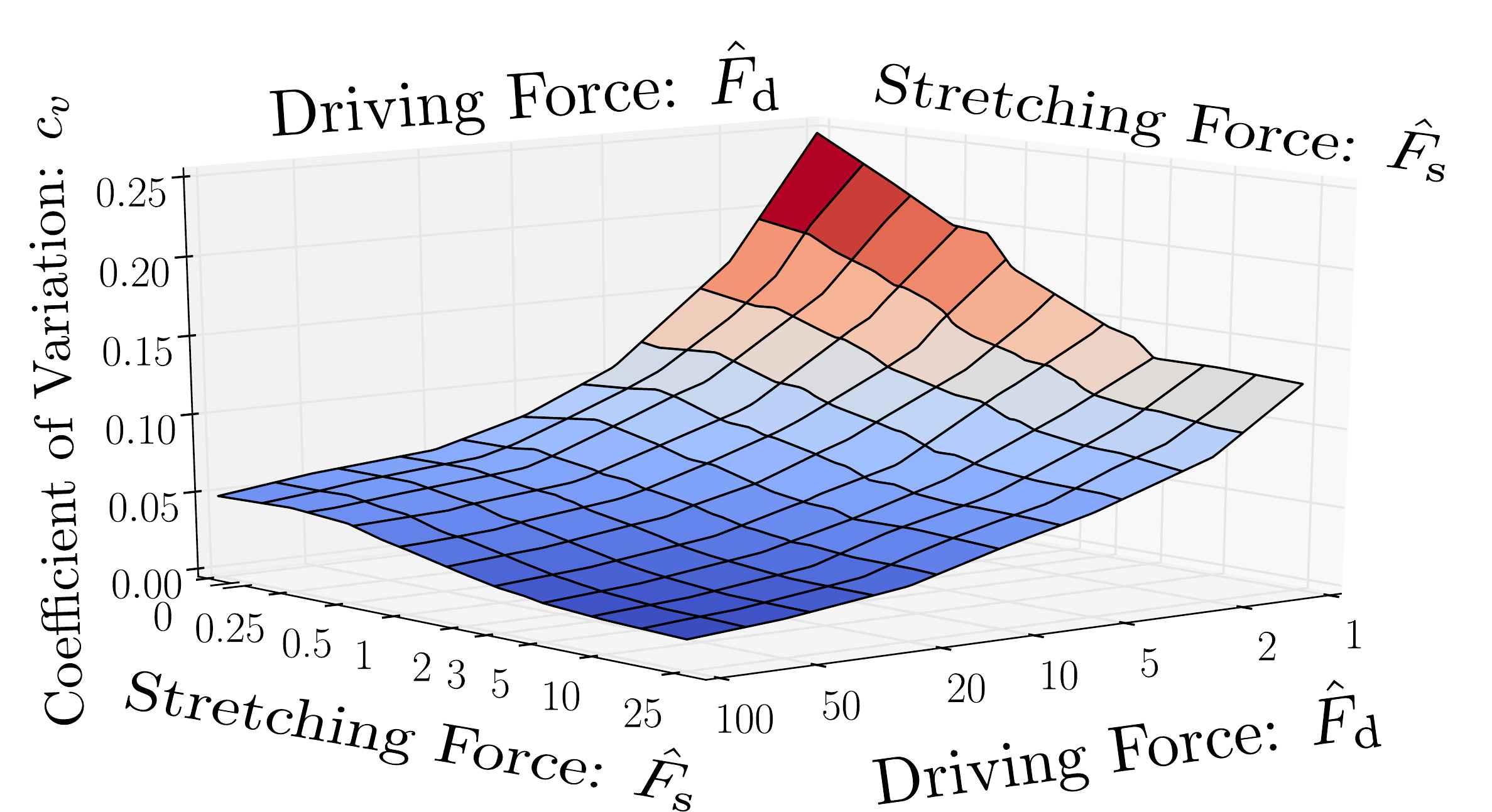} 
	\caption{(Color online) Phase diagram of the coefficient of variation 
$c_v$ as a function of the two variables that control the two noise sources: 
the stretching force $\Fshat$ and the driving force $\Fdhat$. Note the nonlinear 
layout of both force axes.}
	\label{fig:phase}
\end{figure}

Figure~\ref{fig:phase} displays this same data as a phase diagram in $\Fshat$ and $\Fdhat$.
One can see that both increasing the effective driving force (by increasing $\Fd$ or decreasing the temperature)
and pre-stretching the polymer yield improved (lower) coefficients of variation.
The largest effect is observed when both of these factors are employed simultaneously.
However, the region near the lowest $c_v$ is quite flat indicating that both effects saturate
and thus nearly optimal results are obtained once a sufficient strong stretching force is coupled with a sufficiently large P\'{e}clet number.

\section{Conclusions}

In this manuscript, we constructed a simulation model to study translocation such that two sources of noise (conformational and Brownian) can be modulated via two independent control parameters.
This was achieved by a stretching-pulling force scenario in which a force of $\Fs$ is applied to the first monomer and $-\Fs$ is applied to the last monomer to stretch out the polymer both before translocation begins.
These two forces also keep the polymer stretched during translocation while an additional force $\Fd$ is applied to the first monomer which pulls the polymer through pore into the \textit{trans}-region.

We used $\kT$ to control for Brownian noise and $\Fs$ for initial conformations across a range of translocation P\'eclet numbers by varying $\Fd$ and $\kT$.

We found that pre-stretching the polymer has two significant benefits:
not only does the variance of the translocation time $\tau$ decrease, but the mean value of $\tau$ increases.
Hence, the process is slowed down and the accuracy of the measured times is increased; 
both of these are beneficial for DNA sizing  technologies.
The measured benefits are found to strongly depend on the ratio $\Fd \sigma /\kT$ 
since high amounts of diffusion can overwhelm the benefits of limiting the range of initial polymer conformations.

Although the present implementation is difficult to realize experimentally,
these results suggest that stretching the polymer prior to and during translocation via some mechanism could be 
very beneficial for nanopore-based sorting and sequencing applications.
Current work focuses on experimentally viable methods of pre-stretching DNA
that preserve these beneficial effects.

\section{Acknowledgements}

Simulations were performed using the ESPResSo package \cite{limbach2006} on the SHARCNET computer system (www.sharcnet.ca) using VMD \cite{hump1996} for visualization.
This work was funded by NSERC (RGPIN/046434-2013 ) and the University of Ottawa.
HWdH gratefully acknowledges funding from NSERC via the Discovery Grant 2014-06091.
\bibliography{bothEnds}

\end{document}